\documentclass[twocolumn,showpacs,preprintnumbers,amsmath,amssymb]{revtex4}
\usepackage{color}
 
\def\setR{\mathbb{R}}

\newcommand{\ket}[1]{\mid\!#1\,\rangle}

\newcommand{\dron}[3]{\frac{\partial^{#1} {#2}}{\partial{#3}^{#1}}}

\newcommand{\sss}[1]{\scriptscriptstyle #1}

\begin{document}

\title{Conformally covariant quantization of Maxwell field in de Sitter space}

\author{
S. Faci$^1$, E. Huguet$^1$, J. Queva$^1$, J. Renaud$^2$
}
\affiliation{$1$ - Universit\'e Paris Diderot-Paris 7, APC-Astroparticule et Cosmologie (UMR-CNRS 7164), 
Batiment Condorcet, 10 rue Alice Domon et L\'eonie Duquet, F-75205 Paris Cedex 13, France.  \\
$2$ - Universit\'e Paris-Est, APC-Astroparticule et Cosmologie (UMR-CNRS 7164), 
Batiment Condorcet, 10 rue Alice Domon et L\'eonie Duquet, F-75205 Paris Cedex 13, France.
} 
\email{faci@apc.univ-paris7.fr, huguet@apc.univ-paris7.fr, queva@apc.univ-paris7.fr, jacques.renaud@univ-mlv.fr}

\date{\today}

\begin{abstract}
In this article, we quantize the Maxwell (``massless spin one") 
de Sitter field in a conformally invariant gauge.
This quantization is  invariant under the SO$_0(2,4)$ group and consequently under the de Sitter 
group. We obtain a new de Sitter-invariant two-point function which is very simple. Our method relies on the
one hand, on a geometrical point of view which uses 
the realization of Minkowski, de Sitter and anti-de Sitter 
spaces as intersections of the null cone in $\setR^6$ and a moving plane, and on the other hand, on
a canonical quantization scheme of the Gupta-Bleuler type.

\end{abstract}

\maketitle

\section{Introduction}

The main result of the present work is an SO$_0(2,4)$-invariant canonical quantization of the 
Maxwell (``massless spin-one") field in de Sitter space. Precisely, we quantize the
one-form field $A^{\sss H}_\mu$ which fulfills the de Sitter Maxwell equations together with a 
conformal gauge 
condition:
\begin{equation}\label{eqintroMES}
\left \{
\begin{aligned}  
&\square_{\sss H} A_\mu^{\sss H} - \nabla_\mu \nabla A^{\sss H} + 3H^2 A_\mu^{\sss H} = 0\\
&(\square_{\sss H} + 2 H^2) \nabla A^{\sss H} = 0 ,
\end{aligned}
\right .
\end{equation} 
where $12 H^2 = R$, $R$ being the Ricci scalar and $\square_{\sss H}$ the usual Laplace-Beltrami 
operator. As a result, we obtain the following de Sitter-invariant two-point function
\begin{equation}\label{eqintro2pts}
D^{\sss H}_{\mu\nu'}(p,p') = \frac{H^2}{8\pi^2} \left(\frac{1}{{\mathcal Z}-1} g_{\mu\nu'} - 
  n_\mu n_{\nu'}\right) ,
\end{equation} 
where $g_{\mu\nu'}(p,p')$ is the parallel propagator, ${\mathcal Z}$, $ n_\mu$,  $n_{\nu'}$ 
being related to the geodesic distance between the two points $p$ and $p'$ of the de Sitter space 
(see hereafter for more precise statements). This function is simpler than the one obtained
by Allen and Jacobson \cite{AllenJacobson} and, more recently, by Behroozi {\it et al.} \cite{BRTT} 
and Garidi {\it et al.} \cite{GGRT} in ambient $\setR^5$ formalism and, 
Tsamis and Woodard \cite{TsamisWoodard} using the massless limit on the Proca-de Sitter equation.
The reader may also refer to Higuchi and Cheong \cite{HiguchiCheong}
for a recent contribution on the properties of covariant de Sitter two-point functions. 
All these works have been done in the Lorenz gauge ($\nabla A^{\sss H} = 0$).
The simpler form of (\ref{eqintro2pts}) is obtained thanks to a choice of the gauge condition which allows us to preserve the SO$_0(2,4)$-invariance.

To obtain these results we extend the geometrical method used in \cite{pconf1, pconf2} for the scalar field. 
The core of this method is to exploit the realization of Minkowski, de Sitter and anti-de Sitter 
spaces as intersections of the null cone in $\setR^6$ and a moving plane. A continuous change in the position 
of the plane  leads to a continuous transition between spaces. Indeed, the spaces 
are also realized as subsets of the same underlying set (the cone up to the dilations) on which their 
metric tensors are related through a (local) Weyl rescaling. 
This geometric construction allows us, in particular, to easily control the zero-curvature behavior 
of various objects (functions, group generators, ...) and, in the case of Minkowski and de Sitter 
space, to define a common Cauchy surface for field equations. Note that two distinct but related 
notions of ``conformal invariance" are used here: the invariance under  Weyl rescaling and  the 
invariance under the conformal group SO$_0(2,4)$. This point has already been discussed in the case of the scalar  field in \cite{pconf2} and we keep this terminological distinction 
hereafter (see also Kastrup \cite{kastrup} for a review on conformal invariance). 

A second ingredient of our work is the quantization scheme.
The difficulty in maintaining the manifest covariance during the quantization of a gauge invariant
theory is well known. It can be summarized in saying that, in this case,  the canonical quantization scheme fails to give the two-point (Wightman) function  which would be a (causal) reproducing kernel 
for the modes, says $\{\phi_k\}$, solutions of the field equation: 
\begin{equation}\label{eqintroexemple}
\langle {\cal W}(x,\cdot ),\phi_k \rangle = \phi_k (x),
\end{equation}
for any mode $\phi_k$. 
The reason for that is the following: the pure gauge solutions (for instance, the fields $\partial_\mu\Lambda$, in the Minkowski Maxwell case) are known to be orthogonal to any  modes including themselves, so replacing $\phi_k$ by a pure gauge modes in (\ref{eqintroexemple}) should make vanishing the left hand side and not the right hand side. This is impossible. Concerning the canonical quantization of Maxwell field in Lorenz gauge on Minkowski space, one can overcome this difficulty 
by quantizing a field which satisfies, in place of the Maxwell equation 
($\Box A_\mu-\partial_\mu\partial A=0$) together with the Lorenz gauge ($\partial A =0$), a 
covariant but less restrictive equation, 
namely: $\Box A_\mu=0$. The space of solutions of this
equation contains, as a Poincar\'e invariant subset, the solutions of the Maxwell equation in the Lorenz 
gauge. It contains also additional modes (not solution of the Maxwell equations), not orthogonal to 
the pure gauge modes, which solve the above problem. The resultant quantum 
field satisfies the Maxwell equation only in the mean. This
is essentially the Gupta-Bleuler \cite{Gupta, Bleuler} quantization. 

In this paper, we proceed in an analogous way and obtain a conformal quantum field on the de 
Sitter space satisfying (\ref{eqintroMES}) in the mean. Note that, contrary to the 
Maxwell equations,  the Lorenz gauge is not invariant under SO$_0(2,4)$. In Minkowski space the 
use of such a gauge prevents
an SO$_0(2,4)$-invariant quantization of electromagnetism. This problem has been
overcome in the $80'$ \cite{Bayenetal, StokovStoyanov}. The gauge condition used there reduces 
for the free field to $\Box \partial A = 0$. This condition, which can be recognized as the 
Eastwood-Singer gauge \cite{EastwoodSinger} for null curvature, is not SO$_0(2,4)$-invariant alone,  
but the pair Maxwell equations plus Eastwood-Singer condition is. In order to quantize the Maxwell 
field in that gauge, a modified version of the  Gupta-Bleuler formalism, reminiscent of that 
of Nakanishi \cite{Nakanishi}, is used. In it, the whole system, Maxwell equations and gauge 
condition, are replaced by another system containing additional auxiliary fields. 
These fields are then quantized, one of them is in fact used to express a constraint which 
allows us, at the classical level, to recover the Maxwell equations together with the conformal 
gauge condition, and at the quantum level, to determine the subset of physical states. 

In order to generalize this process to de Sitter space, 
we proceed in close analogy with \cite{Bayenetal} by using the well known Dirac's six-cone 
formalism \cite{Dirac6cone, MS} as a starting point for the determination of the auxiliary 
fields. In our system of equations, the application of the constraint leads to the de Sitter Maxwell 
equations together with a covariant gauge (\ref{covgauge}). This system is shown to be 
equivalent to (\ref{eqintroMES}).

Let us remark finally that other quantization schemes are possible, in particular one can 
formulate the classical solutions to the Maxwell equations as gauge equivalent classes 
and then quantize the equivalence classes (see \cite{Dimock} for details). 

Our paper is organized as follows. The geometrical apparatus is introduced in Sec. II, Sec. III 
is concerned by classical field equations.  Sec. IV gives SO$_0(2,4)$ action on the fields, 
Sec. V is devoted to quantization. Some concluding remarks are made in Sec VI. 
 Some formulas and additional points
about Weyl transformation and quantization, and definitions of geometric two-point objects in de Sitter space,  are given in appendices.

\subsection*{Conventions and notations}
Here are the conventions:
\begin{eqnarray*}
\alpha, \beta, \gamma, \delta, \ldots &=&0, \ldots, 5,\\
\mu,\nu,\rho,\sigma,\kappa\ldots &=&  0, \ldots, 3,\\
i, j, k, l, \ldots &=& 1, \ldots, 3.
\end{eqnarray*} 
The indices and superscripts $I, J$ stand for the set $\{c,\mu,+\}$, for instance 
$\{A_{\sss I}\} = \{A_c,A_\mu,A_+\}$. 
The coefficients of the metric $\mbox{diag}(+,-,-,-,-,+)$ of $\setR^{6}$ are denoted $\tilde{\eta}_{\alpha \beta}$:
\begin{equation}
\tilde{\eta}_{55}=\tilde{\eta}_{00}=1=-\tilde{\eta}_{ii}=-\tilde{\eta}_{44}.
\end{equation}
For convenience we set $\eta_{\mu\nu} := \tilde{\eta}_{\mu\nu}$. 
Partial derivatives with respect to the variables $\{y^\alpha\}$ of $\setR^{6}$ are denoted by 
$\tilde{\partial}_\alpha$.

Various spaces and maps are used throughout this paper. Except otherwise stated, quantities related to $\setR^6$ and its null cone $\mathcal{C}$ are labeled with a tilde, those defined on $X_{\sss H}$ (see Sec. \ref{geom} hereafter) are denoted with a super or subscript $H$ except when $H$ takes the null value (Minkowski space) in which case the super or subscript $0$ is omitted. The quantum operator associated with a classical quantity $Q$ is denoted with a hat: $\widehat{Q}$. 

For convenience and readability,  we also
specialize our notations to the de Sitter space 
(the Minkowski space being the particular case where $H = 0$). At a classical 
level our results apply to the anti-de 
Sitter space as well. Expressions relevant for that space can be obtained directly 
from the substitution $H^2 \rightarrow -H^2$.

\section{Geometry and some tools}

\subsection{The spaces}\label{geom}

We first consider the geometrical objects,  namely the spaces and 
how they are related. This part has already been considered 
in \cite{pconf1}; here we want to
complement it, paying a particular attention to its coordinate-free nature.

We begin with realizing the de Sitter, anti-de Sitter, and Minkowski spaces as 
sub-manifolds of $\setR^6$ depending on $H$. The space $\setR^6$, is provided with 
the natural orthogonal coordinates $\{y^\alpha\}$  
and the metric $\tilde{\eta}_{\alpha \beta} = \mathrm{diag}(+,-,-,-,-,+)$.
The five dimensional null cone  $\mathcal{C}$ of $\setR^6$
\begin{equation}\label{C}
\mathcal{C} = \left\{y \in \setR^{6} :  (y^0)^2 - \boldsymbol{y}^2 - (y^4)^2 +(y^5)^2 =0\right\}, 
\end{equation}
is a geometrical object invariant under the action of the conformal group SO$_0(2,4)$.
Let us also define the moving plane
\begin{equation}\label{PH}
P_{\sss H} = \left\{y \in \setR^{6} : (1 + H^2) y^{5} + (1 - H^2) y^{4} = 2 \right\}. 
\end{equation}
The manifold $X_{\sss H} := \mathcal{C} \cap P_{\sss H}$, together with the metric inherited from 
the metric of $\setR^6$, can be shown to be a realization of
the Minkowski ($H = 0$), de~Sitter ($H \neq 0$) or anti-de Sitter (with $H^2$ $\rightarrow$ $-H^2$) 
space. This is also true for the Lie  algebra of generators, naturally parameterized 
by $H$, which reduces to that of Poincar\'e group, SO$(1,4)$ or
SO$(2,3)$ according to the values of $H$ \cite{pconf1}. 

At this point, different values of $H$ correspond to different $X_{\sss H}$ manifolds which are all different sub-manifolds of the cone $\mathcal{C}$. In fact, they can also be viewed as the same manifold with different
$H$-dependent metrics related by a $H$-dependent Weyl factor $K^{\sss H}$. To this end we introduce 
the cone up to the dilations $\mathcal{C'}$, which is the set of the  half-lines of ${\mathcal C}$. 
The realization of $X_{\sss H}$ as a subset of $\mathcal{C'}$ endowed with a $H$-dependent metric has been discussed in \cite{pconf1} with the help 
of a convenient coordinate system. Here we give a coordinate-free presentation. 

We remark that $\mathcal{C}$ has a natural structure of bundle with base $\mathcal{C'}$ and fiber $\setR^+$. The sub-manifold $X_{\sss H}$, for 
a given value of $H$, appears as a partial section of this bundle. This is only a partial section because the natural projection is not onto. This projection allows us to realize the $X_{\sss H}$ 
as subsets of $\mathcal{C'}$. These subsets are endowed with $H$-dependent metrics $g$ which 
are related through a (local) Weyl rescaling:
\begin{equation*}
g_{\mu\nu} = \left(K^{\sss H}\right)^2\eta_{\mu\nu},
\end{equation*}
$K^{\sss H}$ being the Weyl factor. Thus, the de~Sitter, Minkowski and anti-de Sitter spaces 
are realized as subsets of ${\mathcal C}'$. Note that, thanks to the linearity of the action 
of SO$_0(2,4)$, there is a natural action of this group on ${\mathcal C}'$ and hence on $X_{\sss H}$. 
We have proved in \cite{pconf1} that this action is the geometrical one on the de~Sitter, 
Minkowski and anti-de~Sitter spaces.

\subsection{Homogeneous fields}\label{homogeneity}

In this section, we explicitly show the one-to-one correspondence between functions on 
the cone $\cal C$ of $\setR^6$ with a fixed degree of homogeneity, and functions on the de Sitter space $X_{\sss H}$ viewed as a subset of ${\cal C}'$. 
Let us note that the degree of homogeneity of an homogeneous function $f$ is the real number $r$ such that $f(\lambda p) = \lambda^r f(p)$, where $\lambda \in \setR \setminus \{0\}$  and $p\in\setR^6$.

Let us first consider  some hyper-surface of $\setR^6$
defined by some equation $f_{\sss H}(p) = c$, $p\in\setR^6,\ c\in \setR \setminus \{0\}$. In addition, let us assume that $f_{\sss H}$ is homogeneous of degree $1$. Let $p$ be a point of the cone, we note $p^{\sss H}$ the intersection of the hyper-surface with the half line linking $p$ to the origin of $\setR^6$. One can verify that
 $p^{\sss H}=cp/f_{\sss H}(p)$ since
\begin{equation*}
f_{\sss H}\left(\frac{cp}{f_{\sss H}(p)}\right) = \frac{cf_{\sss H}(p)}{f_{\sss H}(p)} = c.
\end{equation*}

For any $p\in\cal C$ we note $[p]$ the corresponding element of ${\cal C}'$: $[p]=\{\lambda p, \lambda>0\}$.
Then, for any  function $\widetilde{F}$, homogeneous of degree $r$, on $\cal C$, we define $\pi_{\sss H}(\widetilde{F})$ on ${\cal C}'$ through
\begin{equation}
\pi_{\sss H}(\widetilde{F})([p])=\widetilde{F}(p^{\sss H}).
\end{equation}
In the following we shorten, as often as possible, this notation to  
$\pi_{\sss H}(\widetilde{F})=F^{\sss H}$, we obtain the useful formula
\begin{equation}
F^{\sss H}([p])=\pi_{\sss H}(\widetilde{F})([p])=\left(\frac{c}{f_{\sss H}(p)}
\right)^r\widetilde{F}(p).
\end{equation}
One can of course recover $\widetilde{F}$ from $F^{\sss H}$ through
\begin{equation}\label{Htotilde}
\widetilde{F}(p)=\left(\frac{f_{\sss H}(p)}{c}\right)^r F^{\sss H}([p]).
\end{equation}
This correspondence allows us to transport different objects such as field equations or group representations from the cone to the de Sitter space, and, as a consequence between
the $X_{\sss H}$ with different values of $H$ (including $H=0$). 

Note that for a 
given $\widetilde{F}$ the corresponding $F^{\sss H}$, which is defined on ${\cal C'}$, is 
not necessarily an intrinsic de Sitter field. Nevertheless we will commit the abuse of language of calling them field all the same.

\section{The field equations on $X_{\sss H}$}
We consider the SO$_0(2,4)$-invariant wave equation for a one-form field in $\setR^6$ \cite{Dirac6cone, MS}
\begin{equation}\label{eqa}
\square_6 \tilde a_\alpha = 0,
\end{equation}
where $\square_6 := \tilde{\eta}^{\alpha\beta}\tilde{\partial}_\alpha\tilde{\partial}_\beta$
and  $\tilde a = \tilde a_\alpha dy^\alpha$ is a one-form field in $\setR^6$ that we choose homogeneous of degree $-1$.  This choice, as shown by Dirac \cite{Dirac6cone}, allows us to consider the field and the equation on the cone $\cal C$ as well. 

In this section we derive a system of equations on $X_{\sss H}$ whose set of solutions contains, as a subset, the SO$_0(2,4)$-invariant solutions of 
the Maxwell equations together with a gauge condition. 

\subsection{A coordinate system}
For practical calculations we use a generalization of the coordinate system used in \cite{Bayenetal},
namely
\begin{equation}\label{coord+muc}
\left \{
 \begin{array}{lcl}
  x^c &=& ~ \dfrac{y_\alpha y^\alpha}{(y^4 + y^5)^2} \\
  x^\mu &=& 2  \dfrac{y^\mu}{y^4 + y^5}\\
  x^+_{\sss H} &=& ~{\displaystyle (1 - H^2) y^4 + (1 + H^2)y^5} .
 \end{array}
\right.
\end{equation}  
In this system, the restriction to the cone $\mathcal{C}$ is expressed by the constraint $x^c = 0$ and 
the restriction to the manifold $X_{\sss{H}}$ by the additional constraint $x^+_{\sss H} = 2$. 
Hence, the coordinate $x^+_{\sss H}$ is nothing but the function $f_{\sss H}$ of 
Sec. \ref{homogeneity}
defining here the moving plane $P_{\sss H}$.
The above system can be inverted in
\begin{equation}\label{coordy}
\left \{
\begin{array}{lcl}
  y^5 &=&  \dfrac{1}{2} \widetilde{K}~x^+_{\sss H}\left(1 
 +x^c - \dfrac{1}{4} \eta_{\mu\nu}x^\mu x^\nu\right) \\
  y^4 &=& \dfrac{1}{2} \widetilde{K}~x^+_{\sss H}\left(1 
 -x^c+ \dfrac{1}{4} \eta_{\mu\nu}x^\mu x^\nu\right) \\
    y^\mu &=&  \dfrac{1}{2} \widetilde{K}~x^+_{\sss H} x^\mu ,
 \end{array}
\right .
\end{equation} 
where 
\begin{equation}\label{K6}
\widetilde{K} := \frac{1}{1 + H^2\left(x^c - \frac{1}{4} \eta_{\mu\nu}x^\mu x^\nu\right)}.
\end{equation}
In the coordinate system $\{x^{\sss I}\}$, 
the homogeneity is carried by the coordinate $x^+_{\sss H}$ alone. This is apparent on the 
expression of the dilation operator: 
\begin{equation}\label{x+dx+}
y\tilde{\partial} = x^+_{\sss H} \dron{}{~}{x^+_{\sss H}}.
\end{equation}

The considerations of Sec. \ref{homogeneity} apply here. Let $[p]=\{\lambda p,\lambda>0\}$ be a point of ${\cal C}'$. All the elements of $[p]$ have the same $\{x^\mu\}$ coordinates (while $x^+_{\sss H}$ depends on $\lambda$).
The system of coordinates $\{x^\mu\}$ thus appears as a coordinate system 
on ${\cal C'}$ and becomes a common system of coordinates for both Minkowski and de Sitter spaces. This system is the so-called polyspherical coordinates \cite{kastrup} on $X_{\sss H}$ which reduces to the cartesian system of coordinates on Minkowski space.

For a given function $\widetilde{F}$, homogeneous of degree $r$,  
on ${\mathcal C}$, one has
\begin{equation}\label{tildetoH}
{F^{\sss H}}(x^\mu) = \left(\frac{x^+_{\sss H}}{2}\right)^{-r}
\widetilde{F}(x).
\end{equation}

One can, for instance, apply this correspondence to the function $\widetilde{K}$ defined in (\ref{K6}), which is homogeneous of degree zero since it does not depend on $x^+_{\sss H}$. One obtains
\begin{equation}\label{K6K}
K^{\sss H} = \frac{1}{1 - \frac{H^2}{4} \eta_{\mu\nu}x^\mu x^\nu}.
\end{equation}
In addition, a direct calculation of the metric shows that
this function is the Weyl factor considered in Sec. \ref{geom}: 
\begin{equation}\label{etag}
g_{\mu\nu} = \left(K^{\sss H}\right)^2 \eta_{\mu\nu}.
\end{equation}
Note also that, for a given  point  $\{x^\mu\}$ on $X_{\sss H}$, the coordinates $\{y^\mu\}$ 
of the corresponding  point of $\setR^6$
 depends on $H$; namely, one has  
$y^\mu = K^{\sss H} x^\mu$.

\subsection{The fields and the extended Weyl transformation}\label{SecFieldandWeyl}

We now introduce the fields $\widetilde{A}_{\sss I}$ which are defined, up to a slight modification on the $dx^+$ component, through the decomposition of the one-form field $\tilde a_\alpha$ on the basis $\{dx\}$: 
\begin{equation}\label{dvpAha}
\tilde a_\alpha dy^\alpha = \widetilde{A}_c dx^c +\widetilde{A}_\mu dx^\mu + \frac{\widetilde{A}_+}{x^+_{\sss H}} dx^+_{\sss H}.
\end{equation}
The $\widetilde{A}_{\sss I}$ being homogeneous, we can define the fields $\{A_{\sss I}^{\sss H}\}$, $I\in\{c, \mu, +\}$. The fields $A_{+}^{\sss H}$ and $A_{c}^{\sss H}$ will be auxiliary fields and the field $A_{\mu}^{\sss H}$ will be, up to the condition  $A_{+}^{\sss H}=0$, the Maxwell field on the de Sitter space. In this case, the $A_{\mu}^{\sss H}$ will be, of course, an intrinsic tensor field on de Sitter space.

Now, expressing the basis $\{dy\}$ in the left hand side in terms of the basis $\{dx\}$ and identifying 
both sides, one obtains the expression of the $\widetilde{A}$ as functions of the $\tilde a$. 
They are homogeneous functions and we can apply 
the correspondence of Sec. \ref{homogeneity}. One obtains
\begin{equation}\label{A(a)}
\left \{
 \begin{array}{lcl}
 {\displaystyle  A_c^{\sss H}} &=& {\displaystyle\left(K^{\sss H}\right)^2
\biggl\{a^{\sss H}_5 \left(1 - H^2\right) }\\
&-& {\displaystyle a^{\sss H}_4\left(1 + H^2\right) - H^2 a^{\sss H}_\mu x^\mu\biggr\}} \\
   {\displaystyle A_\mu^{\sss H}} &=& {\displaystyle \frac{\left(K^{\sss H}\right)^2 }{2}
\biggl\{\Bigl(a^{\sss H}_4 \left(1+H^2\right)-a^{\sss H}_5 \left(1-H^2\right)\Bigr) \eta_{\mu\nu} x^\nu}\\
&+&{\displaystyle  \left . H^2 a^{\sss H}_\sigma x^\sigma  \eta_{\mu\nu} x^\nu + \frac{2}{K^{\sss H}} a^{\sss H}_\mu \right\}} \\
   {\displaystyle A_+^{\sss H}} &=& {\displaystyle 
 K^{\sss H} \left\{a^{\sss H}_5\left(1  - \frac{1}{4} \eta_{\mu\nu}x^\mu x^\nu\right)
 \right. } \\
&+& 
{\displaystyle \left . a^{\sss H}_4\left(1 + \frac{1}{4} \eta_{\mu\nu}x^\mu x^\nu\right) 
+ a^{\sss H}_\mu x^\mu\right\}}  .
 \end{array}
\right .
\end{equation} 
This system can be inverted in
\begin{equation}\label{a(A)}
\left \{
 \begin{array}{lcl}
{\displaystyle a^{\sss H}_5} &=& {\displaystyle \frac{1}{2K^{\sss H}}
\biggl\{A_c^{\sss H} \left(1 - \frac{1}{4} \eta_{\mu\nu}x^\mu x^\nu\right) }\\
&-&{\displaystyle   A_\sigma^{\sss H}x^\sigma +  A_+^{\sss H} K^{\sss H} \left(1 + H^2\right) \biggr\}} \\
{\displaystyle a^{\sss H}_4} &=& {\displaystyle \frac{1}{2K^{\sss H}}
\biggl\{A_c^{\sss H} \left(-1  - \frac{1}{4} \eta_{\mu\nu}x^\mu x^\nu\right) }~~~~~~~~~~~~~~~~~~~\\
&-&{\displaystyle   A_\sigma^{\sss H}x^\sigma +  A_+^{\sss H} K^{\sss H} \left(1 - H^2\right) \biggr\}} \\
 {\displaystyle a^{\sss H}_\mu} &=& {\displaystyle \frac{1}{2K^{\sss H}}
\left\{A_c^{\sss H} \eta_{\mu\nu} x^\nu + 2 A_\mu^{\sss H} \right\}}  .
 \end{array}
\right.
\end{equation}  

We can apply the considerations of the previous section to the field $\tilde a$. Repeated use of formula (\ref{tildetoH}), with $H=H$ and $H=0$, furnishes a relation between $a^{\sss H}$ and $a$, namely
\begin{equation}\label{0toH}
a^{\sss H} =\left(K^{\sss H}\right)^{-1} a.
\end{equation}
Let us remind the reader of our convention which consist in omitting the super or subscript $H$ when $H=0$.  
This formula (\ref{0toH}) together with those linking $A^{\sss H}$ and $a^{\sss H}$
gives the following correspondence, that we  call 
extended Weyl transformation, between the  de Sitter fields  and the Minkowski fields (some of its properties are considered in appendix \ref{WeylEx}):
\begin{equation}\label{extendedWeyl}
\begin{cases}
&{\displaystyle A^{\sss H}_c = A_c - H^2K^{\sss H} A_+}\\
&{\displaystyle A^{\sss H}_\mu = A_\mu + \frac{1}{2} \eta_{\mu\nu} x^\nu} H^2K^{\sss H} A_+\\ 
&{\displaystyle A^{\sss H}_+ = A_+}.
\end{cases}
\end{equation}
We will prove in the following that, for $\tilde{a}$ solution of (\ref{eqa}), $A^{\sss H}_\mu$ can be interpreted as the Maxwell field on the de Sitter space (respectively $A_\mu$ can be interpreted as the Maxwell field on the Minkowski space) up to the condition $A_+=0$. In this case the above extended Weyl transformation becomes  the identity which is, for 
the $A^{\sss H}_\mu$, the ordinary Weyl transformation between one-forms.

\subsection{Equations on $X_{\sss H}$}

The equations for $\{A_{\sss I}^{\sss H}\}$ on
$X_{\sss H}$ are derived from the equation (\ref{eqa}) which is, in some sense, restricted on $X_{\sss H}$. 
This leads to the SO$_0(2,4)$-invariant form of Sec. \ref{eqsnc}. 
A manifestly covariant form is then obtained in Sec. \ref{eqsc}.

\subsubsection{Equations inherited from $\setR^6$}\label{eqsnc}

We first express the operator $\square_6$ in the system $\{x^{\sss I}\}$. Then, using the homogeneity of the one-form field $\tilde a_\alpha$ ($r = -1$) on which the operator acts, and  
applying the constraint $x^c=0$, one obtains the following expression for $\square_6$:

\begin{equation}
\square_6\bigg\vert_{\substack{x^c =\, 0\\ r = -1}}
 = \left(\frac{2}{x^+_{\sss H}}\right)^2 
\left(\frac{1}{\left(K^{\sss H}\right)^2} \partial^2 + \frac{H^2}{K^{\sss H}} x^\mu\partial_\mu + 2 H^2\right),
\end{equation}
where $\partial^2 := \eta^{\mu\nu}\partial_\mu \partial_\nu$. 
As a consequence, the field $a^{\sss H}_\alpha$ satisfies
\begin{equation}\label{boxonXH}
\left(\frac{1}{\left(K^{\sss H}\right)^2} \partial^2 + \frac{H^2}{K^{\sss H}} 
x^\mu\partial_\mu + 2H^2\right)  a^{\sss H}_\alpha(x^\mu) = 0.
\end{equation}
In fact, the above operator can be written in term of 
the Laplace-Beltrami operator on $X_{\sss H}$
acting on a scalar:
\begin{align*}
\square_{\sss H} \phi(x^\mu) &= g^{\mu\nu} \nabla_\mu \nabla_\nu \phi(x^\mu)\\
&= \left(\frac{1}{\left(K^{\sss H}\right)^2} \partial^2 + \frac{H^2}{K^{\sss H}} 
x^\mu\partial_\mu \right) \phi(x^\mu),
\end{align*}
where $\phi$ is a scalar field. Thus (\ref{boxonXH}) reads 
\begin{equation}\label{Box+H2aH}
\left(\square_{\sss H}^s + 2 H^2\right)a^{\sss H}_\alpha=0,
\end{equation}
where $\square_{\sss H}^s$ means that each component  $a^{\sss H}_\alpha$ must be considered as a
scalar. Indeed, the above expression shows that each component of $a^{\sss H}$ satisfies the
equation of a conformal scalar field on $X_{\sss H}$.

Now, using (\ref{a(A)}) in (\ref{boxonXH}) one obtains, after some algebra,
\begin{equation}\label{syst1}
\left \{
\begin{aligned}
&\partial^2 A_\mu^{\sss H} + \partial_\mu A_c^{\sss H} = -\frac{1}{2} \eta_{\mu\nu} x^\nu
 \partial^2 A_c^{\sss H} \\
&\eta^{\sigma\nu}\partial_\sigma A_\nu^{\sss H} +  A_c^{\sss H} = \frac{1}{2} \times\\
&\times (\partial^2 + H^2 K^{\sss H} x^\sigma \partial_\sigma + 2H^2\left(K^{\sss H}\right)^2) A_+^{\sss H} \\
&\partial^2 A_c^{\sss H} = -K^{\sss H} H^2  (\partial^2 + H^2 K^{\sss H} x^\sigma \partial_\sigma+ 2H^2\left(K^{\sss H}\right)^2) A_+^{\sss H}.
\end{aligned}
\right .
\end{equation}  
These equations are the generalization on $X_{\sss H}$ of the system obtained 
in \cite{Bayenetal} in the Minkowskian case. That case is recovered (with a 
slight difference in notations with \cite{Bayenetal}) by setting $H=0$ in (\ref{syst1}) 
which reduces to
\begin{equation}\label{syst1M}
\left \{
\begin{aligned}
&\square A_\mu + \partial_\mu A_c =0 \\
&\square  A_+ - 2 \partial A - 2 A_c = 0  \\
&\square A_c = 0,
\end{aligned}
\right .
\end{equation}  
in which $\partial^2=\square$ because we are on Minkowski space in cartesian coordinates. 
The condition $A_+=0$ {in (\ref{syst1M})},
which is  SO$_0(2,4)$-invariant since $A_+=y^\alpha a_\alpha $, 
allows us to write  $A_c =-\partial A $. The system (\ref{syst1M}) then 
leads to the Maxwell equations and the conformal gauge condition on Minkowski space.

Although not apparent, (\ref{syst1}) is by construction
invariant under the SO$_0(2,4)$ transformations. We claim that
the SO$_0(2,4)$-invariant condition, $A_+^{\sss H}=0$, applied to (\ref{syst1}) gives 
the Maxwell equations and the conformal gauge condition on $X_{\sss H}$; in our particular system of
coordinates this reads
\begin{equation}\label{MaxwellH1}
\left \{
\begin{aligned}
&\partial^2 A_\mu^{\sss H} - \partial_\mu \partial  A^{\sss H}=  0\\
&\partial^2 \partial  A^{\sss H}= 0.
\end{aligned}
\right .
\end{equation}
Here we have set $\partial  A^{\sss H} = \eta^{\sigma\kappa}\partial_\sigma A_\kappa^{\sss H}$, 
in order to make apparent the Minkowskian form of these equations on $X_{\sss H}$, altough 
$\partial  A^{\sss H}$ is not a divergence on $X_{\sss H}$.
Let us stress on the fact that, despite of their Minkowskian form, the above equations are the 
Maxwell equations and the conformal gauge condition on de~Sitter space, although this may not 
be evident. This is due to the use of a specific system of coordinates which makes 
apparent the similarity with the flat case.
We do insist on the fact that this system is SO$_0(2,4)$-invariant on de~Sitter space, because  it is 
nothing but (\ref{eqa}) written in a particular system of coordinates. 
The next section is devoted to writing equations (\ref{syst1}) and (\ref{MaxwellH1}) in a covariant 
form which allows us to recognize the 
Maxwell equations on de~Sitter space.

\subsubsection{Covariant form}\label{eqsc}

In order to find a covariant form of 
(\ref{syst1}) we rewrite all the
operators in (\ref{syst1})
in term of the covariant derivative and the 
connection 
symbols related to the metric $g$. 
Note that, in order to remove explicit references to $x^\mu$, one can use the relation
\begin{equation}\label{xK}
x^\mu = \frac{2}{H^2}\left( K^{\sss H}\right)^{-2} \eta^{\mu\nu}\partial_\nu K^{\sss H} = \frac{2}{H^2} g^
{\mu\nu}\nabla_\nu K^{\sss H}.
\end{equation} 
After some algebra, one obtains 
\begin{equation}\label{syst2}
\left \{
\begin{aligned}  
&\square_{\sss H} A_\mu^{\sss H} - \nabla_\mu \nabla A^{\sss H} + 3H^2 A_\mu^{\sss H} = -\frac{1}{2} \times\\
&\times \nabla_\mu \left(\square_{\sss H} + 2 H^2\right) A_+^{\sss H}\\
&(\nabla - W) A^{\sss H} +
 \left(K^{\sss H}\right)^{-2}A_c^{\sss H} = \frac{1}{2}\left(\square_{\sss H} + 2 H^2\right)A_+^{\sss H}\\
&(\nabla - W )\nabla A_c^{\sss H} = - K^{\sss H} H^2 \left(\square_{\sss H} + 2 H^2\right) A_+^{\sss H},
\end{aligned}
\right .
\end{equation} 
where 
 $\nabla A^{\sss H}$ is the 
divergence of $A^{\sss H}$, and $W$ is the one-form $W:= \mathrm{d} \ln \left(K^{\sss H}\right)^2$ 
of components $W_\mu = \nabla_\mu \ln  \left(K^{\sss H}\right)^2$.  

The previous system (\ref{syst2}) is the covariant form of the system (\ref{syst1}) 
on the manifold $X_{\sss H}$ endowed with the $H$-dependent metric $g$. It is thus a generalization 
to de Sitter and anti-de Sitter (with $H^2 \rightarrow -H^2$) space of the system derived in 
\cite{Bayenetal}. It is worth noting that, owing to equations 
(\ref{A(a)}) and (\ref{Box+H2aH}), (\ref{syst2}) will not have to be solved directly. 

It is now clear that, as for (\ref{syst1}), setting $A_+^{\sss H} = 0$ in (\ref{syst2}) leads to the Maxwell equations and to a 
gauge condition: it is apparent that for $A_+^{\sss H} = 0$ the first line of (\ref{syst2}) are
precisely the Maxwell equations; now, eliminating $ A_c^{\sss H}$ from the remaining two equations
and using the relation $\nabla_\mu  \left(K^{\sss H}\right)^2 =  \left(K^{\sss H}\right)^2 W_\mu$, one obtains the gauge condition
\begin{equation}\label{covgauge}
\nabla^\mu\left(\nabla_\mu + W_\mu \right)\left(\nabla^\nu -
 W^\nu\right)A_\nu^{\sss H} = 0. 
\end{equation} 
Finally, the covariant version of (\ref{MaxwellH1}) reads
\begin{equation}\label{Maxwell2}
\left \{
\begin{aligned}  
&\square_{\sss H} A_\mu^{\sss H} - \nabla_\mu \nabla A^{\sss H} + 3H^2 A_\mu^{\sss H} = 0\\
&\nabla^\mu\left(\nabla_\mu + W_\mu \right)\left(\nabla^\nu - 
 W^\nu\right)A_\nu^{\sss H} = 0.
\end{aligned}
\right .
\end{equation} 
In fact, in relating our gauge condition (\ref{covgauge}) to the Eastwood-Singer gauge \cite{EastwoodSinger}, another covariant system, equivalent to the previous one, will now be obtained.

\subsection{Rewriting the gauge condition}\label{gauges}
We use the notation
\begin{equation*}
D^{\sss H\nu}=\nabla^\mu\left(\nabla_\mu + W_\mu \right)\left(\nabla^\nu -
 W^\nu\right),
\end{equation*}
for the gauge (\ref{covgauge}) which possesses some remarkable properties. 
First, it is invariant under the Weyl transformations between two spaces $X_{\sss H}$. 
This can be derived with the help of formulas for the Weyl transformations (see for 
instance \cite{wald}) and by noting that the conformal weight of $A_\mu^{\sss H}$ is zero. 
One has
\begin{equation}\label{weyltransfgauge}
D^{\sss H_2}A^{\sss H_2} = \left( \frac{K^{\sss H_2}}{K^{\sss H_1}}\right)^{-4}D^{\sss H_1}A^{\sss H_1},
\end{equation}
where $K^{\sss H_1}$ (resp. $K^{\sss H_2}$) is the scalar function  relating the space 
$X_{\sss H_1}$ (resp. $X_{\sss H_2}$) to the Minkowski space (\ref{etag}).

Second,  a straightforward calculation, using (\ref{K6K}), shows that
\begin{equation}
\begin{split}
 D^{\sss H}A^{\sss H} 
 = (\square_{\sss H} &+ 2 H^2) \nabla A^{\sss H} 
 \\
- &  W^\nu 
\left(\square_{\sss H} A_\nu^{\sss H} - \nabla_\nu \nabla A^{\sss H} + 3H^2 A_\nu^{\sss H}\right).
\end{split}
\end{equation}
The system (\ref{Maxwell2}) is then equivalent to 
\begin{equation}\label{Maxwell3}
\left \{
\begin{aligned}  
&\square_{\sss H} A_\mu^{\sss H} - \nabla_\mu \nabla A^{\sss H} + 3H^2 A_\mu^{\sss H} = 0\\
&(\square_{\sss H} + 2 H^2) \nabla A^{\sss H} = 0 .
\end{aligned}
\right .
\end{equation} 
The second line of this system is  the Eastwood-Singer gauge \cite{EastwoodSinger} 
specialized to our constant curvature  space $X_{\sss H}$. This gauge condition is both 
SO$_0(2,4)$-invariant and Weyl invariant between $X_{\sss H}$ spaces only 
on the set of solutions of the Maxwell equations.

The expression (\ref{Maxwell3}) is more compact and more familiar than (\ref{Maxwell2}), nevertheless it is a bit less satisfactory because the Eastwood-Singer gauge condition is not conformally invariant alone.

\section{Action of SO$_0(2,4)$ on the Fields}\label{action}

Now let us turn to the SO$_0(2,4)$ action on fields in connection with the homogeneity. Let us 
consider some tensor field $\widetilde{F}$ of $\setR^6$ defined on $\cal C$ and homogeneous of 
degree $r$. The natural action $\widetilde{T}$ of SO$_0(2,4)$ on $\widetilde{F}$ is
\begin{equation*}
[\widetilde{T}_g\widetilde{F}]_{\sss A'}^{\sss B'} (y) = 
\Lambda_{\sss A'}^{\sss A}(g) \Lambda_{\sss B}^{\sss B'}(g)
\widetilde{F}_{\sss A}^{\sss B}(g^{-1} \cdot y),
\end{equation*}
where $A,B,...$ stands for the indexes of $F$ and $\Lambda_{\sss A'}^{\sss A}$ is a shorthand for the
corresponding product of SO$_0(2,4)$ matrices.
The corresponding action $T^{\sss H}$ of SO$_0(2,4)$ on $F^{\sss H}$ is
defined through the correspondence of Sec. \ref{homogeneity}. 
\begin{equation}
T^{\sss H} := \pi^{\sss H}\widetilde{T}\left(\pi^{\sss H}\right)^{-1}.
\end{equation}
Using the $\{x^{\sss I}\}$ coordinates, we obtain
\begin{equation}\label{actionH}
\begin{split}
[T^{\sss H}_g F^{\sss H}]_{\sss A'}^{\sss B'}(x^\mu) =  
\Lambda_{\sss A'}^{\sss A}(g) \Lambda_{\sss B}^{\sss B'}(g)
\left(\frac{(g^{-1}\cdot x)^+_{\sss H}}{x^+_{\sss H}}\right)^r\\
\left(F^{\sss H}\right)_{\sss A'}^{\sss B'}((g^{-1}\cdot x)^\mu).
\end{split}
\end{equation}
Note that, the expression $(g^{-1}\cdot x)^\mu$ means the component  $\mu$ of the action of $g^{-1}$ on the point of $\setR^6$ of coordinates  $x$, which is nothing but the geometrical action of SO$_0(2,4)$ on $X_{\sss H}$. Moreover, in order to get a more familiar expression for (\ref{actionH}), 
let us consider the invariant square length element of $\setR^6$ restricted on the cone ($x^c=0$), namely
\begin{equation}\label{ds2cone}
ds^2\vert_{x_c=0} = \left(\frac{ x^+_{\sss H}}{2}\right)^2 g_{\mu\nu} dx^\mu dx^\nu.
\end{equation}
The action of SO$_0(2,4)$ on it reads
\begin{align*}
ds^2\vert_{x_c=0} & = d(gs)^2\vert_{x_c=0}
\\
 &= \left(\frac{(g\cdot x)^+_{\sss H}}{2}\right)^2 g_{\mu\nu}( (g\cdot x)^\rho)
d (g\cdot x)^\mu d (g\cdot x)^\nu \\ 
&= \left(\frac{ (g\cdot x)^+_{\sss H}}{2}\right)^2 
\omega_g^2( x^\rho) g_{\mu\nu}( x^\rho) dx^\mu dx^\nu,
\end{align*}
where $\omega_g$ is the scaling term discussed in \cite{pconf2}. Comparing (\ref{ds2cone}) with 
the above expression leads to the identity
\begin{equation}\label{wg-xplus}
\omega_g^2( x^\mu) = \left( \frac{ x^+_{\sss H} }{ (g\cdot x)^+_{\sss H} } \right)^2 ,
\end{equation}
which gives
\begin{equation}
\left(\frac{ (g^{-1}\cdot x)^+_{\sss H} }{ x^+_{\sss H}}\right)^2
= \omega_g^{2}\left(\left(g^{-1}\cdot x\right)^\mu\right).
\end{equation}
Consequently the action (\ref{actionH}) can be rewritten in the more familiar form 
\begin{equation}\label{actionH2}
\begin{split}
[T^{\sss H}_g F^{\sss H}]_{\sss A'}^{\sss B'}(x^\mu) =  
\Lambda_{\sss A'}^{\sss A}(g) \Lambda_{\sss B}^{\sss B'}(g)
\left(\omega_g(g^{-1}\cdot x)\right)^r\\
\left(F^{\sss H}\right)_{\sss A'}^{\sss B'}((g^{-1}\cdot x)^\mu).
\end{split}
\end{equation}
For future reference, 
let us point out that for a scalar field of $\setR^6$, say $\phi$, homogeneous of 
degree $-1$, the action (\ref{actionH2}) becomes
\begin{equation}\label{actionscalaireconf}
\left(T^{\sss H}_g \phi^{\sss H}\right)(x) =
\omega_g^{-1}(g^{-1}\cdot x)
\phi^{\sss H}(g^{-1}\cdot x).
\end{equation}
This is precisely that of a conformal scalar field on de Sitter space.

Now, applying (\ref{actionH2}) to the field $a^{\sss H}_\alpha$ (with $r =-1$), together with the formulas (\ref{A(a)}), (\ref{a(A)}) linking the $a^{\sss H}_\alpha$'s to the $A^{\sss H}_I$'s, one obtains the action of SO$_0(2,4)$ on the fields $A^{\sss H}_I$. The infinitesimal generators follow.
Setting
\begin{align*}
K_\sigma &:= 2 \eta_{\sigma\nu} x^\nu (x\partial) - \eta_{\mu\nu}x^\mu x^\nu\partial_\sigma,\\ 
(M_\sigma)_\nu^\mu &:= 2(\eta_{\sigma\kappa}x^\kappa \delta_\nu^\mu - 
\eta_{\nu\kappa} x^\kappa \delta_\sigma^\mu + x^\mu\eta_{\sigma\nu}),
\end{align*}
the generators read:
\begin{equation*}
\left \{
\begin{aligned}
& \left(K_\sigma^{\sss H}A^{\sss H}\right)_c = (K_\sigma + 4 \eta_{\sigma\nu} x^\nu) A^{\sss H}_c 
+ 4 A^{\sss H}_\sigma \\
&~~~~~~~~~~~~~~+  2H^2\left(K^{\sss H}\right)^2 \eta_{\sigma\nu} x^\nu A^{\sss H}_+~~~~~\\
&\left(K_\sigma^{\sss H}A^{\sss H}\right)_\mu = \left(K_\sigma \delta_\mu^\nu + (M_\sigma)^\nu_\mu\right)A^{\sss H}_\nu \\
&~~~~~~~~~~~~~~- \bigl(2 K^{\sss H} \eta_{\sigma\mu}  + H^2 \left(K^{\sss H}\right)^2 
\eta_{\sigma\nu} x^\nu \eta_{\mu\nu} x^\nu \bigr)A^{\sss H}_+
\\ 
&\left(K_\sigma^{\sss H}A^{\sss H}\right)_+ = K_\sigma A^{\sss H}_+,
\end{aligned}
\right.
\end{equation*}
for the special conformal transformations;
\begin{equation*}
\left \{
\begin{aligned}
&\left(D^{\sss H}A^{\sss H}\right)_c = (x\partial + 2) A^{\sss H}_c +  2H^2 \left(K^{\sss H}\right)^2 
A^{\sss H}_+~~~~~~~~~~~~~~~~~~~~~~~~~~~~~~~\\
&
\left(D^{\sss H}A^{\sss H}\right)_\mu = \left(x\partial + 1\right) A^{\sss H}_\mu - 
H^2 \left(K^{\sss H}\right)^2 \eta_{\mu\nu} x^\nu A^{\sss H}_+\\ 
&
\left(D^{\sss H}A^{\sss H}\right)_+ = x\partial A^{\sss H}_+,
\end{aligned}
\right .
\end{equation*}
for the dilations;

\begin{equation*}
\left \{
\begin{aligned}
& \left(X_{\sigma\epsilon}^{\sss H}A^{\sss H}\right)_c = 
X_{\sigma\epsilon} A^{\sss H}_c~~~~~~~~~~~~~~~~~~~~~~~~~~~~~~~~~~~~~~~~~~~~~~~~~~\\
&
\left(X_{\sigma\epsilon}^{\sss H}A^{\sss H}\right)_\mu = (
X_{\sigma\epsilon}\delta^\nu_\mu + \eta_{\sigma\mu} \delta^\nu_\epsilon -
\eta_{\epsilon\mu}\delta^\nu_\sigma)
A^{\sss H}_\nu\\ 
& 
\left(X_{\sigma\epsilon}^{\sss H}A^{\sss H}\right)_+ = X_{\sigma\epsilon} A^{\sss H}_+,
\end{aligned}
\right.
\end{equation*}
for the rotations, with $X_{\mu\nu} = \eta_{\mu\kappa} x^\kappa \partial_\nu - 
\eta_{\nu\kappa} x^\kappa \partial_\mu$;
 
\begin{equation*}
\left \{
\begin{aligned}
&\left(Y_\sigma^{\sss H}A^{\sss H}\right)_c = (\partial_\sigma - \frac{H}{4}^2 (K_\sigma + 4 
\eta_{\sigma\nu} x^\nu)) A^{\sss H}_c - H^2 A^{\sss H}_\sigma~~~~~~~~~~ \\
&
\left(Y_\sigma^{\sss H}A^{\sss H}\right)_\mu = ( \partial_\sigma\delta^\nu_\mu - \frac{H}{4}^2\left(K_\sigma \delta_\mu^\nu + (M_\sigma)^\nu_\mu\right))A^{\sss H}_\nu \\
& \left(Y_\sigma^{\sss H} A^{\sss H}\right)_+ =(\partial_\sigma - \frac{H}{4}^2 K_\sigma)A^{\sss H}_+,
\end{aligned}
\right .
\end{equation*}
for the other isometries on $X_{\sss H}$, which are given by \cite{pconf2}: 
$Y_\sigma^{\sss H} = P_\sigma^{\sss H} - \frac{H}{4}^2 K_\sigma^{\sss H}$, 
where
\begin{equation*}
\left \{
\begin{aligned}
&\left(P_\sigma^{\sss H}A^{\sss H}\right)_c = \partial_\sigma  A^{\sss H}_c  +
 \frac{1}{2}H^4   \left(K^{\sss H}\right)^2 \eta_{\sigma\nu} x^\nu   A^{\sss H}_+~~~~~\\
&
\left(P_\sigma^{\sss H}A^{\sss H}\right)_\mu = \partial_\sigma\delta^\nu_\mu A^{\sss H}_\nu\\ 
&~~~~~~~~~~~~~- \frac{H^2}{2} K_{\sss H}\left(\eta_{\sigma\mu} + \frac{H^2}{2} K_{\sss H} \eta_{\sigma\nu} x^\nu \eta_{\mu\kappa} x^\kappa \right) A^{\sss H}_+\\
& \left(P_\sigma^{\sss H} A^{\sss H}\right)_+ = \partial_\sigma A^{\sss H}_+ .
\end{aligned}
\right .
\end{equation*}
In view of these results, one can see that, when the physical condition $A^{\sss H}_+ =0$ is fulfilled, the field $A^{\sss H}_\mu$ is an intrinsic de Sitter field.

Finally, let us note that for practical calculation, the finite SO$_0(2,4)$ action on the fields obtained through (\ref{actionH2}) is rather cumbersome. Then, instead of deriving the generators 
directly from it, one can use the extended Weyl transformation  (\ref{extendedWeyl}) as detailed in 
appendix \ref{WeylEx}.

\section{The quantum field}

We now turn to quantum fields. To begin with, we briefly comment on the generic 
Gupta-Bleuler scheme for quantization. Beside undecomposable group representations, 
the mathematical structure underlying this formulation is that of
Krein spaces, which are basically linear spaces endowed with an indefinite scalar product \cite{Bognar}.  
Such a 
structure is known to appear naturally in manifestly covariant canonical quantization of abelian
gauge invariant theory (see for instance \cite{GHR1}).

\subsection{Overview of the Gupta-Bleuler quantization}\label{GBQ}

In order to quantize a tensor field $F$  satisfying some linear equations:
${\cal E}F=0$ on the Minkowski or de Sitter space-time $X_{\sss H}$, one
selects a Hilbert (or Krein) space ${\mathcal K}$ of solutions of the equation equipped with a
scalar product $\langle\ ,\ \rangle$ and carrying a unitary representation of the
symmetry group. 
The only thing to do is to
obtain a causal reproducing kernel ${\cal W}$ for $\cal K$, the Wightman two-point 
function. More precisely, $\cal W$ is a bitensor such that, for each $x\in X_{\sss H}$, ${\cal
W}(x,\cdot): x'\mapsto {\cal W}(x,x')$ is, up to a smearing function on the variable $x$, an
element of $\cal K$ satisfying
\begin{equation}\label{noyau}
\langle {\cal W}(x, \cdot),\psi\rangle = \psi(x),\end{equation}
for any $\psi\in \cal K$, and such that ${\cal W}(x,x')={\cal W}(x',x)$ as soon as $x$ and $x'$ are
causally separated. One can then define the quantum field $\widehat{F}$ through
\begin{equation}\label{lechamp}
\widehat{F}(x)=a\left({\cal W}(x,\cdot)\right)+a^\dag\left({\cal W}(x,\cdot)\right),
\end{equation}
where $a$ and $a^\dag$ are the usual creator and annihilator of the Fock space built onto
$\cal K$.
This field is then a covariant and causal field satisfying the equations ${\cal
E}\widehat{F}=0$ (see Appendix \ref{cq} for a more precise statement and the proof). A way to
obtain an explicit expression for ${\cal W}$ is the following. One considers a family of modes $\{\phi_{k}\}$, that is an Hilbert
(or Krein) basis for $\cal K$,  solution of the field equations 
such that
$\langle\phi_{k},\phi_{k'}\rangle=\zeta_{k}\delta_{kk'}$ where $\zeta_{k}=\pm1$.
Then, the two-point function reads
\begin{equation}\label{deuxpoints}
{\cal W}(x,x')=\sum_k\zeta_k\phi_{k}^*(x)\otimes \phi_{k}(x').
\end{equation}
From this expression, using (\ref{lechamp}) and the anti-linearity and linearity of $a$ and $a^\dag$ respectively, one obtains  the quantum field:
\begin{equation}
\widehat{F}(x)=\sum_k\zeta_k\left(\phi_{k}(x)b_k+\phi_{k}^*(x)b^\dag_k\right),
\end{equation}
where $b_k := a(\phi_{k})$ and $b^\dag_k := a^\dag(\phi_{k})$ are the annihilators and creators of the modes $\phi_k$. The Hilbert space of quantum states $\ket{\psi}$, is then built as usual through the action of the $b^\dag_k$ on the vacuum state of the theory.

As already mentioned in the introduction,
in gauge context, due to the presence of pure gauge solutions, 
such a two-point reproducing kernel does not exist.
In the Gupta-Bleuler scheme, one overcomes this problem by considering an 
enlarged space $\cal H \supset \cal K$ containing some elements not orthogonal 
to the pure gauges.  This space is defined through another equation ${\cal E}'F=0$ 
also invariant under the group. The elements of $\cal K$, called in this context the physical solutions,
satisfy, in addition to the new field equation, a constraint ${\cal G}F = 0$ (for instance 
the Lorenz gauge condition in the usual Gupta-Bleuler quantization of the Maxwell 
field in Minkowski space). 
This classical condition, which allows us to characterize the classical physical solutions, translates
into a quantum condition, which allows us to determine the subspace of physical states (see appendix \ref{cq}).

The new quantum field is of course
covariant and causal, but it  satisfies ${\cal E}'\widehat{F}=0$ instead of ${\cal
E}\widehat{F}=0$.
Nevertheless, one can prove (see the appendix \ref{cq} again) that this last 
equation remains true in the mean for physical
states, precisely:
\begin{equation*}
\langle\psi_1|{\cal E}\widehat{F}|\psi_2\rangle=0,
\end{equation*}
as soon as $|\psi_1\rangle, |\psi_2\rangle$ are physical states.

We now apply this quantization process in our context, 
namely the Maxwell de Sitter field in conformal gauge (\ref{Maxwell3}).
As for the non conformal case, the pure gauge solutions ($A^{\sss H}_\mu = \nabla_\mu\Lambda$, with 
$(\Box_{\sss H} + 2 H^2) \Box_{\sss H} \Lambda = 0$)  are orthogonal to all the solutions including themselves (see Sec. \ref{PS}). As a consequence, 
the space of solutions of (\ref{Maxwell3}) is degenerate and the canonical quantization 
process fails (see above).

Following the Gupta-Bleuler method, we consider the system (\ref{syst2}), instead of (\ref{Maxwell3}), 
for which a causal reproducing kernel can be found.
Thanks to the correspondence between the $A^{\sss H}$ and the $a^{\sss H}$ we need only to solve  
$(\Box_{\sss H}^s+2H^2)a^{\sss H}_\alpha=0$, because it is equivalent to the system (\ref{syst2}). 
In the following, we will define the scalar product, obtain the modes, determine 
the subspace of physical solutions, and then compute the two-point function of the Maxwell field.

\subsection{Scalar product}\label{PS}

Let us define a scalar product on the space of solutions of $(\Box_{\sss H}^s+2H^2)a^{\sss H} = 0$ through
\begin{equation}\label{formPS}
\langle a^{\sss H}, b^{\sss H}\rangle := -\, \widetilde{\eta}^{\alpha \beta}
\langle a^{\sss H}_\alpha, b^{\sss H}_\beta\rangle_s,
\end{equation}
where $\langle, \rangle_s$ is (with a slightly different notation from that used in \cite{pconf1})
the Klein-Gordon scalar product on the space of solutions
of the conformal scalar equation on $X_ {\sss H}$,
\begin{equation}\label{scalarPS}
\langle \phi_1^{\sss H}, \phi_2^{\sss H}\rangle_s := i\, \int_{x^0 = 0}\!\!\!\sigma^\mu\, 
\phi_1^{\sss H *} \stackrel{\leftrightarrow}{\partial_\mu} \phi_2^{\sss H},
\end{equation}
in which $\sigma^\mu$ is the usual surface vector and $\phi_1^{\sss H}$ and $\phi_2^{\sss H}$ denote scalar fields on $X_{\sss H}$. The integral is evaluated on the Cauchy surface of $X_{\sss H}$ defined by $x^0 = 0$.  Implicit summation on repeated indices refers to the metric $g_{\mu\nu}$ on $X_{\sss H}$.

Let us show that the product (\ref{formPS}) is SO$_0(2,4)$-invariant. 
We denote the action defined in (\ref{actionH2}) by 
$T^{{\sss H} f}_g$, for the one-forms, and by $T^{{\sss H} s}_g$, for 
the scalars. Taking into account that the SO$_0(2,4)$ matrix $\Lambda(g)$ appearing in (\ref{actionH2}) depends only of the parameters of the group we have 
\begin{align*}
\langle T^{{\sss H} f}_g a^{\sss H}, T^{{\sss H} f}_g b^{\sss H}\rangle &=\\
- i \widetilde{\eta}^{\gamma \delta}& \int_{x^0 = 0}\!\!\!\!\!\sigma^\mu
\left(\Lambda_\gamma^\alpha \omega_g^{-1} a_\alpha^{\sss H}\right) \stackrel{\leftrightarrow}{\partial_\mu}
\left(\Lambda_\delta^\beta \omega_g^{-1} a_\beta^{\sss H}\right)
\\
&= - \widetilde{\eta}^{\alpha \beta} \langle T^{{\sss H} s}_g a^{\sss H}_\alpha,
 T^{{\sss H} s}_g b^{\sss H}_\beta\rangle_s.
\end{align*}
Thus, the SO$_0(2,4)$ invariance of 
the scalar product between two one-form fields (\ref{formPS}) reduces to the SO$_0(2,4)$ invariance of 
the scalar product (\ref{scalarPS}) between their scalar part. Now, as remarked in the text below equation (\ref{actionH2}) in Sec. \ref{action}, since these parts are homogeneous of degree $-1$ 
they behave as conformal scalars.
Then, $\langle a^{\sss H}_\alpha, b^{\sss H}_\beta\rangle_s$ behaves as the usual Klein-Gordon product
between two conformal scalars, which is known to  
be invariant under SO$_0(2,4)$. Finally, (\ref{formPS}) is SO$_0(2,4)$-invariant.

The product (\ref{formPS}) can be expressed using the $A_{\sss I}^{\sss H}$, one obtains
\begin{equation}\label{PS-A}
\begin{split}
\langle a^{\sss H}, b^{\sss H}\rangle &= - i\, \int_{x^0 = 0}\!\!\!\sigma^\mu\,
\Bigl\{ (K^{\sss H})^2 (
A^{{\sss H} * \nu}\stackrel{\leftrightarrow}{\partial_\mu}B^{\sss H}_\nu)
\\
&+ ( A_\mu^{\sss H *}B_c^{\sss H} - A_c^{\sss H *}B_\mu^{\sss H} ) 
\\
& + \frac{1}{2} ( A_+^{\sss H *}\stackrel{\leftrightarrow}{\partial_\mu}B_c^{\sss H}
+  A_c^{\sss H *}\stackrel{\leftrightarrow}{\partial_\mu}B_+^{\sss H})\\
& + H^2 (K^{\sss H})^2 A_+^{\sss H *}\stackrel{\leftrightarrow}{\partial_\mu}B_+^{\sss H}
\\
& 
+ \frac{1}{2} H^2 K^{\sss H} (A_\mu^{\sss H *}B_+^{\sss H} - A_+^{\sss H *}B_{\mu}^{\sss H})
\Bigr \}.
\end{split}
\end{equation}
For $A^{\sss H}_+ = B^{\sss H}_+ = 0$ the above product reduces to 

\begin{equation}\label{scalarPS-phy}
\begin{split}
\langle a^{\sss H}, b^{\sss H}\rangle &= - i\, \int_{x^0 = 0}\!\!\!\sigma^\mu\,
(K^{\sss H})^2 \Bigl\{ 
A^{{\sss H} * \nu}\stackrel{\leftrightarrow}{\partial_\mu}B^{\sss H}_\nu\\
&- ( A_\mu^{\sss H *} \partial_\nu B^{{\sss H}\nu} -  B_\mu^{\sss H} \partial_\nu A^{{\sss H *}\nu} ) 
\Bigr \}.
\end{split}
\end{equation}

As expected a straightforward calculation shows that the pure gauge solutions, that is written in 
$\{x^\mu\}$ coordinates 
$a = (A^{\sss H}_c = -\partial^2\Lambda$, $A^{\sss H}_\mu = \partial_{\mu}\Lambda$, $A_+^{\sss H} = 0)$ with $ \partial^2\left(\partial^2\Lambda \right)= 0$ are orthogonal 
to all physical states including themselves. In other words, the scalar 
product (\ref{scalarPS-phy}) is gauge invariant.

The product between Minkowskian fields is obtained for $H=0$; it reads
\begin{equation}
\begin{split}
\langle a, b\rangle &= - i\, \int_{x^0 = 0}\!\!\!\sigma^\mu\,
\Bigl\{
A^{*\nu}\stackrel{\leftrightarrow}{\partial_\mu}B_\nu 
+ ( A_\mu^{*}B_c - A_c^*B_\mu ) 
\\
& + \frac{1}{2} ( A_+^*\stackrel{\leftrightarrow}{\partial_\mu}B_c
+  A_c^*\stackrel{\leftrightarrow}{\partial_\mu}B_+)
\Bigr \}.
\end{split}
\end{equation}

\subsection{Modes and physical solutions}\label{solutions}

The equation $(\Box_{\sss H}^s+2H^2)a^{\sss H}_\alpha=0$ is that of a conformal scalar field for each component $a^{\sss H}_\alpha$. As a consequence a set of modes is directly obtained from the solutions of the 
conformal scalar equation. Using the results of \cite{pconf1} the modes on $X_{\sss H}$ reads
\begin{equation}\label{modesXH}
a^{\sss H}_{\sss LM (\gamma)}(x) = \epsilon_{\sss (\gamma)}
\Phi^{\sss H}_{\sss LM}(x),
\end{equation}
where  the one-forms $\epsilon_{\sss(\gamma)}$ are defined trough 
$\epsilon_{\sss (\gamma)\delta} = -\widetilde{\eta}_{\gamma\delta}$ and 
$\Phi^{\sss H}_{\sss LM}(x)$
are the modes which are solutions of the scalar equation $(\Box_{\sss H}+2H^2)\Phi^{\sss H} = 0$ 
(see \cite{pconf1} for details). 
These solutions are normalized with respect to (\ref{formPS}) precisely
\begin{equation} 
\langle a^{\sss H}_{{\sss LM (\gamma)}}, a^{\sss H}_{{\sss L'M' (\delta)}}\rangle = -\tilde\eta_{\gamma\delta}\delta_{\sss LL'}\delta_{\sss MM'}.
\end{equation} 

The general solutions of $(\Box_{\sss H}^s+2H^2)a^{\sss H}_\alpha=0$ are thus given by
\begin{equation}\label{generalsolution-ah}
a^{\sss H}(x)=\sum_{{\sss LM (\gamma)}} b_{{\sss LM (\gamma)}}a^{\sss H}_{{\sss LM (\gamma)}}(x),
\end{equation}
where the $b_{{\sss LM (\gamma)}}$ are some constants with a possible condition of convergence.
Such a solution belongs to the physical subspace of solutions iff the corresponding $A_+^{\sss H}$ vanishes or, equivalently,  in $\{x^\mu\}$ coordinates and using (\ref{A(a)}) iff:
\begin{align}\label{condition}\nonumber
A^{\sss H}_+[a^{\sss H}]&:=(a^{\sss H}_5+a^{\sss H}_4) + \frac{1}{4} \eta_{\mu\nu}x^\mu x^\nu(a^{\sss H}_4-a^{\sss H}_5)
+ a^{\sss H}_\mu x^\mu\\
&\phantom{:}=0.
\end{align}
In order to exhibit a physical solution one can start from a known physical Minkowskian 
solution (for instance a transverse photon $A_\mu$ together with $A_+=0$), then compute $A^{\sss H}$ using (\ref{extendedWeyl}) 
and, finally,  apply the equation (\ref{a(A)}).

\subsection{Two-point functions and quantum fields}\label{2points}
\subsubsection{General form of the two-point function}
We are now looking for the two-point function ${\cal W}^{\sss H}$ satisfying
\begin{equation}
a^{\sss H}(x)=\langle {\cal W}^{\sss H}(x,\ ),a^{\sss H}\rangle.
\end{equation}
This function is obtained through the formula (\ref{deuxpoints}) applied to the above modes:
\begin{equation*}
{\cal W}^{\sss H}=
\sum_{{\sss LM}\gamma}\zeta_\gamma\epsilon_{(\gamma)}
\left(\Phi_{\sss LM}^{\sss H}\right)^*\otimes\epsilon_{(\gamma)}\Phi_{\sss LM}^{\sss H},
\end{equation*}
where $\zeta_\gamma=-\tilde\eta_{\gamma\gamma}$. A straightforward calculation using the results of \cite{pconf1} gives
\begin{equation}\label{generalTwopoints}
{\cal W}^{\sss H}_{\alpha\beta}(x,x')=-\tilde\eta_{\alpha\beta}D^+_{\sss H}(x,x'),
\end{equation}
where $D^+_{\sss H}(x,x')$ is the scalar two-point function.  
For reference, we give here its 
expression in term of the $\{x^\mu\}$ coordinates and as a function of $\mathcal{Z}$ (see appendix \ref{gnn'mu}):

\begin{align}
D^+_{\sss H}(x,x') & = - \frac{1}{4\pi^2} \frac{1}{K^{\sss H}(x) K^{\sss H}(x') 
\eta_{\rho\sigma}(x^\rho - x'^\rho)(x^\sigma - x'^\sigma)} \nonumber
\\
 & = -\frac{H^2}{8\pi^2} \frac{1}{(\mathcal{Z} - 1)}, \label{scalar2pointZ}
\end{align}
in which the regulators are omitted for the sake of readability.

\subsubsection{Quantum field and physical states}

We can now define the quantum field, using (\ref{lechamp}). It reads
\begin{equation}
\widehat{a}^{\sss H}(x)=\sum_{LM\gamma}a^{\sss H}_{\sss LM (\gamma)}(x)
b_{\sss LM (\gamma)}+a^{{\sss H}\,*}_{\sss LM (\gamma)}(x)
b^\dag_{\sss LM (\gamma)},
\end{equation}
$b_{\sss LM (\gamma)}$ and $b^\dag_{\sss LM (\gamma)}$ being the annihilators and creators of the mode $a^{\sss H}_{\sss LM (\gamma)}$.  The quantum field $\widehat{A}^{\sss H}_\mu$, that is the Maxwell de Sitter field, is obtained from the field $\widehat{a}^{\sss H}$ 
through (\ref{A(a)}). 

Before discussing the two-point function of the Maxwell field on de Sitter space, 
let us comment about physical states in relation with field equations. The quantum
states are built, as usual, by applying the  creators $b^\dag_{\sss LM (\gamma)}$
 on the vacuum of the theory: $\ket{0}_{\sss H}$.
The subset of physical states can be formally determined thanks to the classical physical solutions, that is those $a^{\sss H}$ which satisfy (\ref{condition}): $A_+^{\sss H}[a^{\sss H}] = 0$. 
In fact, to define the physical states it is sufficient to say that they are created from physical 
solutions: $\ket{a^{\sss H}}$ is a physical state iff
\begin{equation}\label{physicalstateDef}
\ket{a^{\sss H}} =a^\dag(a^{\sss H}) \ket{0}_{\sss H},
~\mathrm{and}~A_+^{\sss H}[a^{\sss H}] = 0.
\end{equation}
These physical states satisfy the quantum counterpart of (\ref{condition})
\begin{equation}\label{quantumcondition}
\widehat{A}^{\sss H (+)}_+ \ket{a^{\sss H}} = 0,
\end{equation}
where $\widehat{A}^{\sss H (+)}_+ $ is the annihilator part of $\widehat{A}^{\sss H}_+ $. This implies that the equality 
\begin{equation}
\langle a^{\sss H}\mid \widehat{A}_+(x)\mid b^{\sss H}\rangle =0,
\end{equation}
holds as soon as $\ket{a^{\sss H}}$ and $\ket{b^{\sss H}}$ are physical states. That is proved in great generality in the appendix \ref{cq}, but one can verify this directly in our case. 
From the definitions of 
$\widehat{A}^{\sss H (+)}$ and $\ket{a^{\sss H}}$, one obtains
\begin{equation*}
\widehat{A}^{\sss H (+)}_+ \ket{a^{\sss H}} = A^{\sss H}_+[a^{\sss H}] \ket{0}_{\sss H}, 
\end{equation*} 
which is true for all classical solutions $a^{\sss H}$. The right hand side of the above equality is
obviously zero only for a physical solution $a^{\sss H}$ and thus (\ref{quantumcondition}) follows.
As a consequence, for physical states one has
\begin{equation*}
\begin{cases}
&\langle a^{\sss H}\mid   \square_{\sss H} \widehat{A}_\mu^{\sss H} - \nabla_\mu \nabla \widehat{A}^{\sss H} + 3H^2 \widehat{A}_\mu^{\sss H}   \mid b^{\sss H}\rangle=0\\
&\langle a^{\sss H}\mid  (\square_{\sss H} + 2 H^2) \nabla \widehat{A}^{\sss H}  \mid b^{\sss H}\rangle=0.
\end{cases}
\end{equation*}
In other words, the field fulfills the Maxwell equation together with the conformal gauge in the mean on physical states.

All the above considerations are in close analogy with the usual covariant quantization of 
the Maxwell field in Lorenz gauge in Minkowski space. 
The classical condition which corresponds to (\ref{condition}) is the Lorenz gauge condition 
and its quantum counterpart reads $\partial \widehat{A}^{(+)} \ket{0} = 0$. Indeed, the above 
formulation can be transposed to this well known situation as well. Now, an important property of 
physical states in this original Gupta-Bleuler quantization is that their norms are non-negative 
(precisely: positive for transverse photons and null for pure gauges). 
We now show that the same property holds for the physical 
states (\ref{physicalstateDef}) with respect to the scalar product (\ref{scalarPS-phy}). 
Here is the proof. Let us first consider the Minkowskian case ($H=0$), the physical solutions 
belong to the subset of solutions of the Maxwell equations which satisfy the gauge condition 
$\square \partial A = 0$.  This subset contains the Lorenz gauge as a subset. Now, given a solution 
$A_\mu$ which satisfies $\square \partial A = 0$ and $\partial A \neq 0$ one can always find a gauge 
transformation, that is a function $\Lambda$, such that $A'_\mu := A_\mu + \partial_\mu \Lambda$ 
satisfies $\partial A' = 0$. Precisely, $\Lambda$ is solution of  $\square \Lambda = - \partial A$. Now,
$A'_\mu$, which is  a solution of the Maxwell equations in the Lorenz gauge has a non-negative norm. Then, since the scalar product (\ref{scalarPS-phy}) is gauge 
invariant, the same conclusion holds for $A_\mu$. That is, in the Minkowskian case the scalar 
product (\ref{scalarPS-phy}) is non-negative on the subspace of physical solutions. It remains to show that
this conclusion extends to the de Sitterian case. 
Indeed, the map $a \mapsto a^{\sss H}$ defined by (\ref{0toH})
is nothing but that introduced in \cite{pconf1} in the study of
the conformal scalar, and we have already shown that this map is unitary. The conclusion thus 
follows.

\subsubsection{Maxwell de Sitter two-point function}

The Maxwell de Sitter two-point function can now be defined through
\begin{equation}\label{MaxweldS2ptDef}
D^{\sss H}_{\mu\nu'}(x,x')=\,_{\sss H\,}\!\langle0\mid 
\widehat{A}^{\sss H}_\mu(x)\widehat{A}^{\sss H}_{\nu'}(x')
\mid 0\rangle_{\sss H}.
\end{equation}
The quantum field $\widehat{A}^{\sss H}_\mu(x)$ is related to  $\widehat{a}^{\sss H}(x)$ by (\ref{A(a)}).
Since, as usual, the field satisfies
\begin{equation}\label{tpf1}
{\cal W}^{\sss H}(x,x')=\,_{\sss H\,}\!\langle0\mid \widehat{a}^{\sss H}(x)\widehat{a}^{\sss H}(x')\mid0\rangle_{\sss H},
\end{equation}
this allows us, taking (\ref{generalTwopoints}) into account, to compute (\ref{MaxweldS2ptDef}) straightforwardly. After some algebra, it reads in $\{x^\mu\}$ coordinates
\begin{equation}\label{MaxweldS2ptGur}
\begin{split}
&D^{\sss H}_{\mu\nu'}(x,x') = - K^{\sss H}(x) K^{\sss H}(x') 
\biggl[ \eta_{\mu\nu'} + \\
&\frac{~H^2}{2} \eta_{\mu\kappa} \eta_{\rho\nu'} \biggl(
K^{\sss H}(x)  x^{\kappa}(x^{\rho} - x'^{\rho}) +
K^{\sss H}(x') x'^{\rho}(x'^{\kappa} - x^{\kappa}) \\
& +\frac{~H^2}{2} \Bigl(K^{\sss H}(x) K^{\sss H}(x') \frac{1}{2}\eta_{\theta\sigma}(x^\theta - x'^\theta)(x^\sigma - x'^\sigma) \Bigr) \times\\
& x^{\kappa}x'^{\rho}  
 \biggr) \biggr] D^+_{\sss H}(x,x').
\end{split}
\end{equation}
This expression is not really convenient as it stands; note however, that it makes apparent 
that the Minkowskian two-point function (given in \cite{Bayenetal}) is recovered for $H = 0$. 
Now, since $D^{\sss H}_{\mu\nu'}(x,x')$ is a de Sitter invariant function, it must be a function 
of the intrinsic and invariant quantity $\mathcal{Z}$ 
(see appendix \ref{gnn'mu} for details). In fact, using (\ref{un}), (\ref{deux}) and (\ref{trois}), 
the expression (\ref{MaxweldS2ptGur}) can be recast under the form
\begin{equation*}
D^{\sss H}_{\mu\nu'}(x,x') = -  \bigl( g_{\mu\nu'}  - ({\mathcal Z} - 1) n_\mu n_{\nu'} \bigr) D^+_{\sss H}(x,x'),
\end{equation*}
where the geometrical objects $\mathcal{Z}$, $g_{\mu\nu'}$, $n_\mu$ and $n_{\nu'}$ are explicitly 
defined in appendix \ref{gnn'mu}.
Finally, using the explicit form (\ref{scalar2pointZ}) of $D^+_{\sss H}(x,x')$, the one-form 
two-point function rewrites
\begin{equation}\label{MaxweldS2ptFinal}
D^{\sss H}_{\mu\nu'}(x,x') = \frac{H^2}{8\pi^2} \left(\frac{1}{{\mathcal Z}_\varepsilon -1} 
g_{\mu\nu'} - n_\mu n_{\nu'}\right),
\end{equation}
where ${\mathcal Z}_\varepsilon := {\mathcal Z} - i\varepsilon (x^0 - x'^0)$ includes the regulator. 
Note that
there is no other singular point 
than ${\mathcal Z} = 1$. In addition, this two-point function has clearly the Hadamard behavior
and thus our vacuum is the Euclidean one. This behavior could be expected since the 
modes (\ref{modesXH}) are basically inherited from those of the conformal scalar field equation 
on $X_{\sss H}$. These modes are related to their Minkowskian counterpart through a Weyl 
transformation. In this respect, the vacuum in the de Sitter theory 
is in close relation with that of the Minkowskian theory. Since in solving the scalar equation 
in \cite{pconf1} we implicitly choose the usual Minkowski vacuum (that corresponds to positive 
frequency modes) we keep track of this choice in (\ref{MaxweldS2ptFinal}).

The above result differs from that of Allen and Jacobson \cite{AllenJacobson} which is repeated here, 
with our conventions, for
convenience:
\begin{align*}
D^{\sss H \mathrm{(AJ)}}_{\mu\nu'}(x,x') &= \alpha({\mathcal Z}) g_{\mu\nu'} + 
\beta({\mathcal Z}) n_\mu n_{\nu'},
\end{align*}
where
\begin{align*}
        \alpha(\mathcal{Z}) &= 
        \frac{H^2}{24\pi^2}\Bigl[ -\frac{3}{\mathcal{Z}-1}
        + \frac{1}{\mathcal{Z} + 1} \\
        &\qquad+ \Bigl(\frac{2}{\mathcal{Z} + 1} + \frac{2}{(\mathcal{Z} + 1)^2}\Bigr)
        \log\Bigl(\frac{1 - \mathcal{Z}}{2}\Bigr)\Bigr], 
        \\
        \beta(\mathcal{Z}) &=
        \frac{ H^2}{24\pi^2}\Bigl[
                1 -\frac{2}{\mathcal{Z}+1} \\
        &\qquad+ \Bigl(\frac{2}{\mathcal{Z} + 1} + \frac{4}{(\mathcal{Z} + 1)^2}\Bigr)
        \log\Bigl(\frac{1 - \mathcal{Z}}{2}\Bigr)\Bigr].
\end{align*}
It is not surprising that these two-point functions are different since 
different gauges have been used. 
On the contrary, one can consider the gauge invariant quantity 
\begin{equation*}
\,_{\sss H\,}\!\langle0\mid \widehat{F}^{{\sss H} \mu\nu}(x)\widehat{F}_{\mu'\nu'}^{\sss H}(x')\mid0\rangle_{\sss H}=   \nabla^{[\mu}\nabla_{[\mu'} D^{\nu]}_{\nu']}(x,x'),
\end{equation*}
which is the two-point function for the Faraday field strength tensor $F = \mathrm{d} A$. A straightforward calculation shows that we obtain the same result as Allen and Jacobson \cite{AllenJacobson}.

Finally, let us point out a property of our conformal quantization in connection with the 
two-point function obtained by Garidi et al. \cite{GGRT}. Their quantization 
proceeds in close analogy with the usual Gupta-Bleuler quantization in which
the classical lagrangian of the theory is modified by adding a so-called gauge fixing term.  This term corresponds to the Lorenz gauge and is parameterized by a constant $c$. The two-point function 
obtained in \cite{GGRT} (formula 5.29) is the sum of the two-point function 
(\ref{MaxweldS2ptFinal}) and of a term 
which is a non-vanishing function $c$. In other words, no value of of the gauge fixing parameter $c$ 
can lead to the two-point function (\ref{MaxweldS2ptFinal}).

\section{Conclusion}

In order to conclude this work, we would like to stress three facts.

We choose the strategy of preserving as far as possible the SO$_0(2,4)$-symmetry 
of the Maxwell equations during the process of quantization. This led us to take
a gauge condition which could, at first, appear complicated compared to the usual Lorenz condition 
in de Sitter space. In fact, it leads to a simple form of the two-point function.

In writing the de Sitter and Minkowski spaces as subsets of the cone up to the dilations, we can easily 
obtain the limit $H=0$ for all the objects of our paper, including modes and quantum field.

Finally, our construction gives an explicit expression for the quantum fields and the states,
not only for the two-point function.

\appendix
\section{The extended Weyl transformation}\label{WeylEx}

In this appendix, we give some properties of the extended Weyl transformation defined 
in (\ref{extendedWeyl}).
It is convenient for practical calculations to introduce the notation
\begin{equation}\label{notmat}
\mathcal{A}^{\sss H} = 
\begin{pmatrix}
A_c^{\sss H}\\
A_\mu^{\sss H}\\
A_+^{\sss H}
\end{pmatrix}.
\end{equation}
Keeping the usual left product for the matrices, the extended Weyl 
transformation then reads 
\begin{equation*}
\mathcal{A}^{\sss H}= S_{\sss K} \mathcal{A},
\end{equation*}
with
\begin{equation}\label{weylgene}
S_{\sss K} := \begin{pmatrix}
1&0&-H^2K^{\sss H}\\
0&1&\frac{1}{2} W_\mu\\
0&0&1
\end{pmatrix}.
\end{equation}

The form of $S_{\sss K}$ makes obvious the conservation of the condition $A_+ = 0$ under the 
extended Weyl transformation. Moreover, a straightforward computation shows that the system 
(\ref{syst1}) is left invariant in the sense that:
$\{\mathcal{A}_{\sss I}^{\sss H}\}=\{S_{\sss K}(\mathcal{A}_{\sss I})\}$ is solution of (\ref{syst1}) iff $\{\mathcal{A}_{\sss I}\}$ is solution of (\ref{syst1M}). Finally, that transformation  allows us to 
transport not only the
fields but also the operators acting upon them. Explicitly,  one define an operator 
$\widehat{O}^{\sss H}$ from the operator 
$\widehat{O}^0$ by

\begin{equation}\label{Weylop}
\widehat{O}^{\sss H} := S_{\sss K} \widehat{O}^0 S^{-1}_{\sss K}.
\end{equation}

As an application, one can derive the results of Sec. \ref{action} in a convenient way: 
using the matrix notation for the generators, the action SO$_0(2,4)$ on the 
fields $\{A_{\sss I}^{\sss H}\}$ is obtained, thanks to (\ref{Weylop}), 
from that on the fields $\{A_{\sss I}\}$.

\section{Quantization}\label{cq}

In this appendix, we prove the assertions of Sec. \ref{GBQ}. We consider, on 
some space-time,  an Hilbert or Krein space $\cal K$ of functions satisfying some 
equations ${\cal E}\psi(x)=0$. The space $\cal K$ carries a unitary representation 
$U$ of the symmetry  group defined through
\begin {equation}
\left(U_g\psi\right)(x)=M_{g}(x)\psi(g^{-1}\cdot x),
\end{equation}
where $M_{g}$ is a product of real matrices acting on the tensor $\psi$.
We assume the existence of a causal reproducing kernel for $\cal K$ such that
\begin{equation}\label{noyauApp}
\langle {\cal W}(x, \cdot),\psi\rangle = \psi(x),\end{equation}
for any $\psi\in \cal K$, and, moreover, ${\cal W}(x,x')={\cal W}(x',x)$ as soon as $x$ and $x'$ are causally separated. 
Then, one can define a quantum field through
$$\widehat{F} (x)= a({\cal W}(x, \cdot))+a^\dag({\cal W}(x, \cdot)),$$
where $a$ and $a^\dag$ are the usual annihilator and creator on the Fock space built on $\cal K$.
As a result, this field is causal, covariant and satisfies (in the distribution sense), ${\cal E}\widehat{F}=0$. Note that the invariance of ${\cal W}$ is in the consequences, not in the hypothesis.

Let us begin with causality. Using well-known properties of annihilators and creators (see for instance  \cite{grt})
one obtains for $x$ and $x'$ causally separated
\begin{eqnarray*}
[\widehat{F}(x),\widehat{F}(x')]&=&
\langle {\cal W}(x, \cdot),{\cal W}(x', \cdot)\rangle 
-\langle {\cal W}(x', \cdot),{\cal W}(x, \cdot)\rangle\\
&=&{\cal W}(x',x)-{\cal W}(x,x')\\
&=&0.
\end{eqnarray*}
This proves that this field is causal.

The covariance of the field is defined through
\[
\underline{U}_g\widehat{F}(x)\underline{U}_{g^{-1}}=M_{g}(x)\widehat{F}(g^{-1}x),
\]
where $\underline{U}$ is the natural action of the group on the Fock space.
The corner stone of the proof is the following identity that we will now prove:
\[
U_{\check{g}}{\cal W}(x,\check{x}')=
U_{\check{g}^{-1}}{\cal W}(\check{x},x'),
\]
where the $\ \check{}\ $ indicates that  the group acts on the variable $x'$ in the left 
hand side and on the variable $x$ in 
the right hand side. This is due to the formula (\ref{noyauApp}), in fact, for any $\psi\in \cal K$:
\begin{align*}
\langle U_{\check{g}}{\cal W}(x,\check{\cdot}),\psi\rangle
&=\langle {\cal W}(x,\cdot),U_{g^{-1}}\psi\rangle\\
&=\left(U_{g^{-1}}\psi\right)(x)\\
&=M_{g^{-1}}(x)\psi(g x)\\
&=M_{g^{-1}}(x)\langle {\cal W}(g x,\cdot),\psi\rangle\\
&= \langle M_{g^{-1}}(x) {\cal W}(g x,\cdot),\psi\rangle\\
&=\langle U_{\check{g}^{-1}} {\cal W}(\check{x},\cdot),\psi\rangle.
\end{align*}
The covariance follows immediately, using the standard formula
$\underline{U}_ga(\psi)\underline{U}_{g^{-1}}=a(U_g\psi)$.

From the very definition of ${\cal W}$, one can see that $\check{\cal E}{\cal W}(x,\check{x}')=0$.
Moreover, using once again (\ref{noyauApp}), we have also $\check{\cal E}{\cal W}(\check{x},x')=0$, in fact, for any $\psi\in \cal K$:
\begin{align*}
\langle \check{\cal E}{\cal W}(\check{x},\cdot),\psi\rangle
&=\check{\cal E}\langle {\cal W}(\check{x},\cdot),\psi\rangle \\
&={\cal E}\psi(x)\\
&=0.
\end{align*}
The desired equality ${\cal E}\widehat{F}(x)=0$ follows immediately.

Suppose now that we are in gauge context, we get a space $\cal H$ larger than $\cal K$
defined through the equations ${\cal E}'\psi=0$. We assume that $\cal H$ is invariant under the group action. The same process as above can run and we obtain a field which is causal and covariant. But the quantum field obeys to the equations 
${\cal E}'\widehat{F}=0$ and not ${\cal E}\widehat{F}=0$. Nevertheless, the last equation remains true in the mean on physical states:
\begin{equation}
\langle\psi_1|{\cal E}\widehat{F}(x)|\psi_2\rangle=0,\label{inthemean}
\end{equation}
for any physical states $\psi_1,\psi_2$. In order to prove that, we define $\widehat{F}^{(+)}$
annihilator part of $\widehat{F}$ and consider a ``one particle sector" physical state $\ket{\psi}=
a^\dag(\psi)\ket{0}$ where $\psi$ is a physical solution: ${\cal E}\psi=0$. Then
\begin{align*}
{\cal E}\widehat{F}^{(+)}(x)\ket{\psi}&={\cal E}a\left({\cal W}(x,\cdot)\right)\ket{\psi}\\
&={\cal E}a\left({\cal W}(x,\cdot)\right)a^\dag(\psi)\ket{0}\\
&={\cal E}\langle{\cal W}(x,\cdot),\psi\rangle \ket{0}\\
&={\cal E}\psi(x) \ket{0}\\
&=0.
\end{align*}
This can be generalized easily to ``many particles" sectors, the equation  (\ref{inthemean}) follows immediately.

\section{Intrinsic quantities for bitensors in de Sitter space}\label{gnn'mu}

Following Allen and Jacobson \cite{AllenJacobson} (where the reader is referred for proofs and details),
any maximally symmetric bitensor 
(that is, invariant under the isometry group of a maximally symmetric manifold, here
the de Sitter space)
can be decomposed in a unique way as sum of products of fundamental objects. They are: 
the metric at points $p$ and $p'$ of the manifold and three quantities 
related to the length $\mu(p,p')$,
of the geodesic
 from $p$ to $p'$ ($\mu$ being imaginary when the geodesic is spacelike), namely:
\begin{description}
 \item[~~~] $n_{\mu}(p,p') = \nabla_{\mu} \mu(p,p')$ is the unit tangent vector to the geodesic at 
the point $p$,
 \item[~~~] $n_{\nu'}(p,p') = \nabla_{\nu'} \mu(p,p')$  is the unit tangent vector to the geodesic 
at the point $p'$,
 \item[~~~] $g_{\mu\nu'}(p,p') = \dfrac{1}{C} \nabla_{\mu} n_{\nu'}(p,p') - n_{\mu}(p,p')n_{\nu'}(p,p')$ 
is the parallel propagator along the geodesic,
\end{description}
where we use the usual convention that a primed (resp. not primed) index refers to a primed (resp. not primed) point. 
The factor $C$ will be given in what follows.

In order to define the standard variable $\mathcal{Z}$, let us introduce the five-dimensional ``ambient" Minkowski space with metric 
$\bar{\eta} = \mathrm{diag}(+,-,-,-,-)$. We will use small roman letters $a,b,c,...$ to denote indices running from $0$ to $4$. The de Sitter space can be viewed as the 
sub-manifold defined by the equation 
\begin{equation*}
\bar{\eta}_{ab} X^a X^b = -H^{-2}, 
\end{equation*}
where $\{X^a\}$ denotes ambient space cartesian coordinates. A point $p$ on the de Sitter space is 
associated to the vector $X(p)$ of coordinates $X^a(p)$. The ambient coordinates are related to the
coordinates $\{x^\mu\}$ through
\begin{equation}\label{coordAmbiant}
 \begin{cases}
   X^\mu &=K^{\sss H} x^\mu ,\\
   X^4 &= \dfrac{1}{H} \left(2 K^{\sss H} - 1\right).
 \end{cases}
\end{equation} 
The function ${\mathcal Z}={\mathcal Z}(p,p')$ is then defined through
\begin{equation}{\mathcal Z}:= -H^2 \bar{\eta}_{ab}X^aX'^b,\end{equation}
where $X = X(p)$ and $X' = X(p')$.
The geodesic distance 
$\mu(p,p')$ is related to ${\mathcal Z}$ by
\begin{equation}\label{zmu}
\mathcal{Z}= \cosh\left(H \mu \right), \mathcal{Z} 
\geqslant -1.
\end{equation}
The case $\mathcal{Z} < -1$ corresponds to the situation where
$p'$ is lying in the interior of the light cone of the antipodal of $p$ and, in this case, there is no geodesic connecting $p$ and $p'$. Nevertheless,  $\mathcal{Z}$ is always defined and one can define $\mu(p,p')$ through an analytic continuation (see \cite{AllenJacobson} again).
As a function of $\mathcal{Z}$ the factor $C$ reads 
\begin{equation*}
C  =  \frac{-H}{\sqrt{(\mathcal{Z}^2-1)}}.
\end{equation*} 

Now, using (\ref{coordAmbiant}), one has
\begin{align*}
\bar{\eta}_{ab} X^a X'^b &= K^{\sss H} K'^{\sss H} xx' - \frac{1}{H^2} (2  K^{\sss H} - 1 )
(2  K'^{\sss H} - 1 )\\
&= - K^{\sss H} K'^{\sss H} \sigma_0(x,x') - \frac{1}{H^2},
\end{align*} 
where $\sigma_0(x,x') = (\eta_{\mu\nu} (x^\mu - x'^\mu)(x^\nu-x'^\nu))/2$ and $K'^{\sss H}=~K^{\sss H}(x')$. Thus,
in the $\{x^\mu\}$ coordinates $\mathcal{Z}$ reads 
\begin{equation*}
\mathcal{Z} = H^2 K^{\sss H}K'^{\sss H}\sigma_0 + 1,
\end{equation*}
from which one obtains
\begin{equation*}
\nabla_{\mu} \mathcal{Z}(x,x') = H^2 K^{\sss H} \eta_{\mu\kappa}\left(\frac{\mathcal{Z}-1}{2}x^{\kappa} +
K'^{\sss H} (x^\kappa-x'^\kappa) \right).
\end{equation*}
From (\ref{zmu}) we also have
\begin{equation*}
\nabla_{\mu} \mathcal{Z}(x,x') = -\frac{H^2}{C^{~}} n_{\mu}.
\end{equation*}
In our  system of coordinates $\{x^\mu\}$, we find that
\begin{align}\label{un}
n_{\mu} &=   -C K^{\sss H} \eta_{\mu\kappa} 
\left[ \dfrac{\mathcal{Z}-1}{2} x^\kappa + K'^{\sss H}(x^\kappa - x'^\kappa) \right]
\\ \label{deux}
n_{\nu'} &=  - C K'^{\sss H} \eta_{\nu'\kappa}
\left[ \dfrac{\mathcal{Z}-1}{2} x'^\kappa + K^{\sss H}(x'^\kappa - x^\kappa) \right]
\\ 
g_{\mu\nu'} &=   (\mathcal{Z}-1) n_{\mu} n_{\nu'} 
 +  K^{\sss H}K'^{\sss H} 
\left[
\eta_{\mu\nu'} \right. \nonumber \\
&\left.- H^2  \dfrac{\mathcal{Z}-1}{4} \eta_{\mu\kappa} x^\kappa \eta_{\nu'\rho}x^\rho 
+ \dfrac{H^2}{2} \eta_{\mu\kappa}\eta_{\nu'\rho}  \right. \nonumber \\
&\left. \times \left( K^{\sss H}x^\kappa 
(x^\rho-x'^\rho) + K'^{\sss H}x'^\rho (x'^\kappa-x^\kappa) 
\right) 
\right].\label{trois}
\end{align}
Combined with (\ref{MaxweldS2ptGur}), this gives the crucial result (\ref{MaxweldS2ptFinal}).

\end{document}